\documentclass[12pt,draftcls,onecolumn,journal,letterpaper]{IEEEtran}
\usepackage{color}
\usepackage{amsmath}
\usepackage{theorem}
\usepackage{psfig}
\usepackage{epsfig}
\usepackage{amsfonts}
\usepackage{amssymb}
\usepackage{cite}
\usepackage{graphics}
\newtheorem{thm}{Theorem}
\newtheorem{lem}{Lemma}

\renewcommand{\baselinestretch}{1.5}

\begin{document}
\title{Fairness in Multiuser Systems with Polymatroid Capacity
Region\footnote{The material in this paper was presented in part at
the 43th Allerton Conference, Monicello, IL, Sep. 2005 and IEEE
International Symposium on Information Theory (ISIT) , Seattle, WA,
USA, July 2006.}}
\author{Mohammad A. Maddah-Ali, Amin Mobasher, and Amir K.
Khandani\\
Coding \& Signal Transmission Laboratory (www.cst.uwaterloo.ca),\\
Dept. of Elec. and Comp. Eng., University of Waterloo,\\
Waterloo, Ontario, Canada, N2L 3G1,\\
e-mail: \{mohammad, amin, khandani\}@cst.uwaterloo.ca
\thanks{Financial supports provided by Nortel, and the corresponding matching funds by the Federal
government through Natural Sciences and Engineering Research Council
of Canada (NSERC) and Province of Ontario through Ontario Centres of
Excellence (OCE) are gratefully acknowledged.} }
 \maketitle
\markboth{IEEE Transactions on Information Theory (Submitted)}{}
\begin{abstract}
For a wide class of multi-user systems, a subset of capacity region
which includes the corner points and the sum-capacity facet has a
special structure known as polymatroid. Multiaccess channels with
fixed input distributions and multiple-antenna broadcast channels
are examples of such systems. Any interior point of the sum-capacity
facet can be achieved by time-sharing among corner points or by an
alternative method known as {\em rate-splitting}.
 The main purpose of this paper
is to find a point on the sum-capacity facet which satisfies a
notion of fairness among active users.  This problem is addressed in
two cases: (i) where the complexity of achieving interior points is
not feasible, and (ii) where the complexity of achieving interior
points is feasible. For the first case, the corner point for which
the minimum rate of the active users is maximized (max-min corner
point) is desired for signaling. A simple greedy algorithm is
introduced to find the optimum max-min corner point. For the second
case, the polymatroid properties are exploited to locate a
rate-vector on the sum-capacity facet which is optimally fair in the
sense that the minimum rate among all users is maximized (max-min
rate). In the case that the rate of some users can not increase
further (attain the max-min value), the algorithm recursively
maximizes the minimum rate among the rest of the users. It is shown
that the problems of deriving the time-sharing coefficients or
rate-spitting scheme can be solved by decomposing the problem to
some lower-dimensional subproblems. In addition, a fast algorithm to
compute the time-sharing coefficients to attain a general point on
the sum-capacity facet is proposed.
\end{abstract}
\section{Introduction}
In the multi-user scenarios, multiple transmitters/receivers share a
common communication medium, and therefore, there is an inherent
competition in accessing the channel.  Information theoretic results
for such systems imply that in order to achieve a high spectral
efficiency, the users with stronger channel should have a higher
portion of the resources. The drawback to this is the loss of the
fairness among the users.   Providing fairness, while achieving
high-spectral efficiency, is thus a challenging problem.

A lot of research has addressed this problem and suggested different
criteria to design a fair system. One of the first criteria is known
as {\em max-min} measure. In this method, the main effort is to
maximize the minimum rate of the users, by giving the highest
priority to the user with the worst channel.  In other words, this
method penalizes the users with better channel and sacrifices
overall efficiency.

By relaxing the strict condition on fairness, the spectral
efficiency can be increased. By compromising between fairness and
throughput,  proportional fairness is proposed in~\cite{Prop_Fair}.
Based on this criterion, the rates of users with a stronger channel
can be increased with the cost of decreasing the rates of users with
a weaker channel. Any change in the rates is acceptable if the total
proportional increase in the rates of some users is larger than the
total proportional decrease in the rates of the rest.  In fact, by
relaxing the strict condition on fairness, the spectral efficiency
increases.   In~\cite{NASH_BAR}, a criterion based on Nash
Bargaining solution in the context of Game Theory is proposed. This
method generalizes the proportional fairness and increases the
efficiency of the system.

All of the aforementioned methods deal with a general multi-user
system. However, for a wide class of multi-user systems, the
capacity region has a special structure that we can exploit to
provide fairness. Particularly in some multiuser systems, the
boundary of the capacity region includes a facet on which the
sum-rate is maximum (sum-capacity facet). In such systems, one can
benefit from the available degrees of freedom, and determine the
fairest rate-vector on the sum-capacity facet.

As a special case, we consider a class of multi-user systems, in
which  the whole or a subset of the capacity region which includes
the corner points and the sum-capacity facet forms a structure known
as polymatroid. For this class of multi-user systems, the
sum-capacity facet has $a!$ corner points, where $a$ is the number
of users with non-zero power (active users). The sum-capacity facet
is the convex hull of these corner points. This means that the
interior points of the sum-capacity facet can be attained by
time-sharing among such corner points. As an example of such
systems, it is shown that the capacity region of multiaccess
channels (MAC) with fixed and independent input distributions forms
a polymatroid~\cite{ploy_tse}. In MAC, the sum-capacity is achieved
by successive decoding. Applying different orders for the users in
successive decoding results in different rate-vectors, all with the
sum-rate equal to the sum-capacity. The resulting rate-vectors
correspond to the corner points of the sum-capacity facet.  Any
point in the convex hull of these corner points is on the boundary.
In~\cite{Succ_dec}, it is proven that the Marton inner bound
(see~\cite{marton}) for capacity region of the broadcast channel
under fixed joint probability of the auxiliary and input variables,
 with some conditions, has a polymatroid
structure\footnote{Throughout the paper, we deal with the systems
where the underlying capacity region or a its subset which included
sum-capacity facet forms a polymatroid. Apparently, the proposed
method can be applied over any achievable region which has the
similar geometrical structure. In this case, the sum-capacity facet
is replaced with maximum-sum-rate facet.}. As another example, we
will show that a subset of the capacity region for multiple-input
multiple-output (MIMO) broadcast channel which includes the corner
points forms a polymatroid.

In~\cite{ploy_tse}, the optimal dynamic power allocation strategy
for time-varying single-antenna multiple-access channel is
established. To this end, the polymatroid properties of the capacity
region for time-invariant multiple-access channel with fixed input
distributions have been exploited. In~\cite{ShiFri05}, the
polymatroid properties have been used
 to find a  fair power allocation strategy. This problem is
formulated by representing a point on the face of the
contra-polymatroid (see~\cite{ploy_tse,poly_tse2}) as a convex
combination of its extreme points.

This article aims at finding a point on the sum-capacity facet which
satisfies a notion of fairness among active users by exploiting the
properties of polymatroids. In order to provide fairness, the
minimum rate among all users is maximized (max-min rate). In the
case that the rate of some users can not increase further (attain
the max-min value), the algorithm recursively maximizes the minimum
rate among the rest of the users. Since this rate-vector is in the
face of the polymatroid, it can be achieved by time sharing among
the corner points. It is shown that the problem of deriving the
time-sharing coefficients to attain this point can be decomposed to
some lower-dimensional subproblems. An alternative approach to
attain an interior point for multiple access channels is {\em rate
splitting}~\cite{rate_spil1,rate_spil2}. This method is based on
splitting all input sources except one into two parts and treating
each spilt input as two virtual inputs (or two virtual users). By
splitting the sources appropriately and successive decoding of
virtual users in a suitable order, any point on the sum-capacity
facet can be attained~\cite{rate_spil1,rate_spil2}. Similar to the
time-sharing procedure, we show that the problem of rate-splitting
can be decomposed to some lower dimensional subproblems.

There are cases that the complexity of achieving interior points is
not feasible. This motivates us to compute the corner point for
which the minimum rate of the active users is maximized (max-min
corner point). A simple greedy algorithm is introduced to find the
max-min corner point.

The rest of the paper is organized as follows. In Section
\ref{secII}, the structure of the polymatroid is presented. In
addition, the relationship between the capacity region of some
channels and the polymatroid structure is described. Section
\ref{secIII} discusses the case in which the optimal fair corner
point is computed. In Section \ref{secIV}, the optimal fair
rate-vector on the sum-capacity facet is computed by exploiting
polymatroid structures. In addition, it is shown that the problem of
deriving the time-sharing coefficients and rate-splitting can be
solved by decomposing the problem into some lower-dimensional
subproblems.

\textit{Notation:} All boldface letters indicate vectors (lower
case) or matrices (upper case).  $\det(\mathbf{H})$ and
$\mathbf{H}^{\dag}$ denote the determinant and the transpose
conjugate of the matrix $\mathbf{H}$, respectively. $\mathbf{M}
\succeq 0 $ represents that the matrix $\mathbf{M}$ is positive
semi-definite. $\mathbf{1}_n$ represents an $n$ dimensional vector
with all entries equal to one. $E$ is a set of integers
$E=\{1,\cdots,|E|\}$, where $|E|$ denotes the cardinality of the set
$E$. The set function $f: 2^E \longrightarrow \mathcal{R}_{+}$ is a
mapping from all subsets of $E$ (there are a total of $2^{|E|}$
subsets) to the positive real numbers. A permutation of the set $E$
is denoted by $\pi$ and $\pi(i)$, $1\leq i\leq |E|$, represents the
element of the set $E$ located in the $i^{th}$ position after the
permutation. For an $a$-dimensional vector ${\bf x}=\{ x_1,
x_2,\ldots,x_a\} \in \mathcal{R}^a$ and $S\subset E$, ${\bf x}(S)$
denotes $\sum_{i\in S} x_i$. Also, for a set of positive
semi-definite matrices $\mathbf{D}_{i}$, ${\bf D}(S)$ represents
$\sum_{i \in S}\mathbf{D}_{i}$.

\section{Preliminaries}\label{secII}

\subsection{Polymatroid Structure}

\emph{Definition}~\cite[Ch. 18]{Wel76}: Let $E=\{1,2, \ldots, a \}$
and $f:~2^E \longrightarrow \mathcal{R}_{+}$ be a set function. The
polyhedron
\begin{equation}\label{ine:poly}
\mathcal{B}(f,E)=\{(x_1,\ldots,x_{a}): \mathbf{x} (S)\leq f(S),
\forall S \subset E, \forall x_i\geq 0 \}
\end{equation}
is a polymatroid, if the set function $f$ satisfies
\begin{eqnarray}
\label{poly:norm}(normalized)\ &  f( \emptyset )=0 & \\
\label{poly:inc} (increasing)\ \ & f(S)\leq f(T) \ \textrm{if} \  S \subset T & \\
\label{poly:subm}(submodular)\ & f(S)+f(T) \geq f( S \cap T)+f(S\cup
T) &
\end{eqnarray}
Any function $f$ that satisfies the above properties is termed as
{\em rank function}. Note that (\ref{ine:poly}) imposes $2^{|E|}$
constraints on any given vector $(x_1,\ldots,x_{a}) \in
\mathcal{B}(f,E)$.

 Corresponding to
each permutation  $\pi$ of the set $E$, the polymatroid
$\mathcal{B}(f,E)$ has a corner point ${\bf v}(\pi) \in
\mathcal{R}^{a}_{+}$ which is equal to:
\begin{eqnarray} \label{eq:v_pi}
v_{\pi(i)}(\pi)= \left \{
\begin{array}{ll}
f(\{\pi(i)\}) & \ i=1  \\
\; & \; \\
f(\{\pi(1),\ldots,\pi(i)\}) & \; \\
\;\;\;\;\;\; -f(\{\pi(1),\ldots,\pi(i-1)\}) & \ i=2,\ldots,a
\end{array}
\right .
\end{eqnarray}
Consequently, the polymatroid $\mathcal{B}(f,E)$ has $a!$ corner
points corresponding to different permutations of the set $E$. All
the corner points are on the facet $\mathbf{x}(E)=f(E)$. In
addition, any point in the polymatroid on the facet
$\mathbf{x}(E)=f(E)$ is in the convex hull of these corner points.
The hyperplane $\mathbf{x}(E)=f(E)$ is called as dominant face, or
simply face of the polymatroid. In this paper, we use the term {\em
sum-capacity facet} to denote the face of the polymatroid.

\subsection{Capacity Region and Polymatroid Structure}
For a wide class of multi-user systems, the whole or a subset of the
capacity region forms a polymatroid structure. As the first example,
consider a multiaccess system with $a$ users, where the distribution
of inputs are independent and equal to $p(x_1), \ldots, p(x_M)$.
Then, the capacity region of such a system is characterized
by~\cite{Ahlswede,Liao}
\begin{equation}
\left\{\mathbf{r}\in \mathcal{R}^{a}_+ | \mathbf{r}(S) \leq
I\left(y;\{x_i, i\in S\}|\{x_i, i\in S^c\}\right) \quad \forall S
\subset E
 \right\},
\end{equation}
where $y$ is the received signal, $\mathbf{r}$ represents rate
vector, $I$ denotes the mutual information, and $S^c$ is equal to
$E-S$. It has been shown that the above polyhedron forms a
polymatroid~\cite{ploy_tse}.

As the second example, we consider the capacity region of a
multiple-antenna broadcast system. In the sequel, we  show that a
subset of the capacity region which includes the corner points and
sum-capacity facet forms a polymatroid.

Consider a MIMO Broadcast Channel (MIMO-BC) with $M$ transmit
antennas and $K$ users, where the $k^{\textrm{th}}$ user is equipped
with $N_k$ receive antennas. In a flat fading environment, the
baseband model of this system is given by
\begin{equation}\label{model}
\mathbf{y}_k=\mathbf{H}_k\mathbf{s}+\mathbf{w}_k, \quad 1 \leq k\leq
K,
\end{equation}
where $\mathbf{H}_k \in \mathcal{C}^{N_k\times M}$ denotes the
channel matrix from the base station to user $k$, $\mathbf{s} \in
\mathcal{C}^{M\times 1}$ represents the transmitted vector, and
$\mathbf{y}_k \in \mathcal{C}^{N_k\times 1}$ signifies the received
vector by  user $k$. The vector $\mathbf{w}_k \in
\mathcal{C}^{N_k\times 1}$ is a white Gaussian noise with zero-mean
and identity-matrix covariance. Consider an  order of the users
$(\pi(1), \pi(2),\ldots, \pi(K))$. By assuming that user $\pi(i)$
knows the codewords selected for the users $\pi(j)$, $j=1,\ldots,
i-1$, the interference of the users $\pi(j)$, $j=1,\ldots, i-1$,
over user $\pi(i)$ can be effectively canceled based on
dirty-paper-coding theorem~\cite{Costa}. Therefore, the rate of user
$\pi(i), i=1,\ldots,K,$ is equal to
\begin{equation}\label{rate:BC}
r_{\pi(i)}=\log \frac{\det \left (
\mathbf{I}_{N_k,N_k}+\mathbf{H}_{\pi(i)} \left (\sum_{j \geq i}
\mathbf{Q}_{\pi(j)}\right )\mathbf{H}^{\dag}_{\pi(i)} \right
)}{\det\left( \mathbf{I}_{N_k,N_k}+\mathbf{H}_{\pi(i)} \left
(\sum_{j
> i} \mathbf{Q}_{\pi(j)}\right )\mathbf{H}^{\dag}_{\pi(i)} \right
)},
\end{equation}
where $\mathbf{Q}_{\pi(j)}$ is the covariance of the signal vector
to user $\pi(j)$. The capacity region is characterized as the convex
hull of the union of such rate-vectors over all permutations
$(\pi(1), \pi(2),\ldots, \pi(K))$ and over all positive
semi-definite covariance matrices $\mathbf{Q}_i$, $i=1,\ldots, K$
such that $\mbox{Tr}\left ( {\sum_{i=k}^{K} \mathbf{Q}_i} \right )
\leq P_T$, where $P_T$ denotes the total transmit
power~\cite{Br_Cap_RG_J}.
In~\cite{BroadCap_Caire,BroadCap_Gold,BroadCap_Tse}, a duality
between the MIMO-BC and the MIMO-MAC is established. In the dual
MIMO-MAC, the channel between user $k$ and the base station is
$\mathbf{H}^{\dag}_{k}$ and the covariance of the power allocated to
user $k$ is $\mathbf{P}_k$. The relationship between
$\mathbf{P}_{k}$ and $\mathbf{Q}_{k}$, $k=1,\ldots,K$, has been
derived~\cite{BroadCap_Gold}. The duality is used to characterize
the sum-capacity of the MIMO-BC as follows
\begin{eqnarray}\label{eq:sum}
\nonumber r_{\textrm{SC}} & = & \max_
{\mathbf{P}_1,\ldots,\mathbf{P}_K} \log \det \left(
\mathbf{I}_{M,M}+\displaystyle \sum_{k=1}^K
\mathbf{H}^{\dag}_k\mathbf{P}_k\mathbf{H}_k \right ). \\ \nonumber
& s.t. & \sum_{k=1}^K{\mbox{Tr}(\mathbf{P}_k)} \leq P_T, \\
& \ & \mathbf{P}_k\succeq 0
\end{eqnarray}
The above optimization problem determines the power  allocated to
each user in the dual MIMO-MAC, and consequently, the power of each
user in the MIMO-BC. Note that only a subset of users is active and
the power allocated to the rest is zero. Equation (\ref{eq:sum})
determines the so-called sum-capacity facet. If the cardinality of
the set of active users is $a$, i.e. $E=\{1,\cdots,a\}$, the
sum-capacity facet has $a!$ corner points corresponding to different
permutations of the active users. Note that the rates of the
non-active users remain zero regardless of the permutation. The
corner point corresponding to a permutation can be computed using
(\ref{rate:BC}). Assuming the active users are indexed by
$i=1,\ldots, a$, we define
\begin{equation}\label{eq:Di}
\mathbf{D}_i=\mathbf{H}^{\dag}_i\mathbf{P}^*_i\mathbf{H}_i,
i=1,\ldots,a,
\end{equation}
where $\mathbf{P}^*_i$, $i=1,\ldots,a$, correspond to optimizing
matrices in (\ref{eq:sum}). It is shown that the corner point in
(\ref{rate:BC}) can be reformulated as~\cite{BroadCap_Gold}
\begin{equation}\label{rate:MAC}
r_{\pi(i)}=\log \frac{ \det \left(\mathbf{I}_{M,M}+\sum_{j \leq i}
\mathbf{D}_{\pi(j)} \right )}{\det \left(\mathbf{I}_{M,M}+\sum_{j <
i} \mathbf{D}_{\pi(j)}  \right )}, \quad i=1,\ldots,a,
\end{equation}
which is the corner point of the dual MAC.

Regarding the polymatroid structure of the multiaccess channels and
considering the duality of the MIMO-MAC and MIMO-BC, we can observe
the polymatroid structure of a subset of MIMO-BC capacity region
which includes the sum-capacity facet. However, to provide a better
insight about the problem, we introduce a special polymatroid and
establish its relationship with the capacity region of the MIMO-BC.
For a set of positive semi-definite matrices $\mathbf{D}_{i}$, we
define the set function $g$ as,
\begin{equation}\label{eq:gS}
g(S)=\log \det \left(\mathbf{I}+{\bf D}(S) \right) \quad
\textrm{for} \quad S \subset E.
\end{equation}
\begin{lem}
Given $g(S)$ defined in (\ref{eq:gS}), the polyhedron
$\mathcal{B}(g,E)$ defined as follows is a polymatroid.
\begin{equation}
\mathcal{B}(g,E)=\{(x_1,\ldots,x_{a})\in
\mathcal{R}^a_{+}:\mathbf{x}(S)\leq g(S), \; \forall S \subset E\}.
\end{equation}
\end{lem}
\begin{proof}
Clearly, $g(\emptyset)=0$. Assume $\mathbf{B}\succeq 0$ and
$\mathbf{C} \succeq 0$ are two Hermitian matrices. If
$\mathbf{B}-\mathbf{C} \succeq 0$, then $\det(\mathbf{B}) \geq
\det(\mathbf{C})$ \cite[Proposition I.2]{Br_Cap_RG_J}. Furthermore,
if $\mathbf{\Delta}\succeq 0$, then \cite[Proposition
I.3]{Br_Cap_RG_J}
\begin{equation}\label{ineq:pos}
\frac{\det (\mathbf{\Delta}+\mathbf{B}+\mathbf{C})}{\det (
\mathbf{\Delta}+\mathbf{B})} \leq \frac{\det
(\mathbf{B}+\mathbf{C})}{\det(\mathbf{B})}.
\end{equation}
Using above properties, it is straight-forward to prove
(\ref{poly:inc}) and (\ref{poly:subm})  for the set function $g(.)$.
\end{proof}

\begin{figure}[bhpt]
\centerline{ \psfig{figure=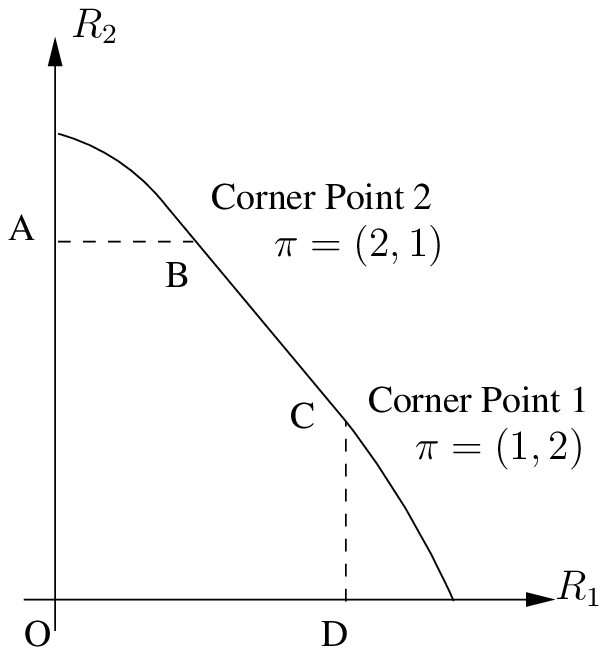,width=2.1 in,height=2.1 in}}
\caption{ \small{Capacity region of the MIMO-BC and its corner
Points. The region OABCD is a polymatroid. The line BC is the
sum-capacity facet.}} \label{capacity}
\end{figure}

In the set function $g(S)$, define ${\bf D}_i$ as defined in
(\ref{eq:Di}). It is easy to verify that the polymatroid
$\mathcal{B}(g,E)$ is a subset of the capacity region of the
MIMO-BC. The hyperplane $\mathbf{x}(E)=g(E)$ and its corner points
(\ref{rate:MAC}) are the same as the sum-capacity facet and its
corner points. Due to this property, we focus on the polymatroid
$\mathcal{B}(g,E)$ (see Fig. \ref{capacity}).

\section{The Fairest Corner Point}\label{secIII}
As mentioned, in some cases, the complexity of computing and
implementing an appropriate time-sharing or rate-splitting algorithm
is not feasible. This motivates us to compute the corner point for
which the minimum rate of the active users is maximized (max-min
corner point). In the following, we present a simple greedy
algorithm to find the max-min corner point of a general polymatroid
$\mathcal{B}(f, E)$.

\textbf{Algorithm I}
\begin{enumerate}
\item Set $\alpha =a$, $S= \varnothing$.
\item Set $\pi^*(\alpha)$ as
\begin{equation}\label{eq:1}
\pi^*(\alpha)= \arg \min_{z \in E , z \notin \mathcal{S}}
f\left(E-S-\{z\}\right).
\end{equation}
\item If $\alpha >1$, then  $S
\longleftarrow S \cup \{ \pi^*(\alpha) \}$, $\alpha\longleftarrow
\alpha-1$, and go to Step 2; otherwise stop.
\end{enumerate}
The following theorem proves the optimality of the above algorithm.
\begin{thm}\label{thm:max-min}
Let the vector $\mathbf{v}(\pi^*)$ be the corner point of the
polymatroid $\mathcal{B}(f, E)$ corresponding to the permutation
$\pi^*=(\pi^*(1),\ldots,\pi^*(a))$. For any other permutation
$\pi=(\pi(1),\ldots,\pi(a))$,
\begin{equation}\label{ineq:max_min}
\min_{i}v_{\pi^*(i)}(\pi^*) \geq \min_i v_{\pi(i)}(\pi).
\end{equation}
\end{thm}

\begin{proof}
Assume that in the permutation $\pi^*$, the user $\theta$ which is
located in position $l$ in the permutation $\pi^*$ ( i.e.
$\theta=\pi^*(l)$) has the minimum rate
\begin{equation}
v_{\pi^*(l)}(\pi^*)=\min_{i}v_{\pi^*(i)}(\pi^*).
\end{equation}
 Let us define two sets:

\begin{itemize}
\item The set of users located  before $\pi^*(l)$ in $ \pi^*$:
$\Phi=\{\pi^*(1),\ldots, \pi^*(l-1)\}$.

\item The set of users located  after $\pi^*(l)$ in $\pi^*$:
$\Psi=\{\pi^*(l+1),\ldots, \pi^*(a)\}$.
\end{itemize}
Using (\ref{eq:v_pi}), we have
\begin{equation}\label{eq:rate_pil}
v_{\theta}(\pi^*)=f(\Phi \cup \{\theta \}) - f(\Phi).
\end{equation}
In the following, we consider different scenarios which generate new
permutations and prove that in all cases, (\ref{ineq:max_min}) is
valid.

\textbf{Case 1.} \emph{Permutation in $\Phi$ and $\Psi$:}  By
considering (\ref{eq:rate_pil}), it is apparent that any permutation
of the users in $\Phi$ and $\Psi$ does not change the rate of the
user $\pi^*(l)$ (see Fig. \ref{fig:case1}).
\begin{figure}[tbhp]
\centering
\includegraphics[scale=1]{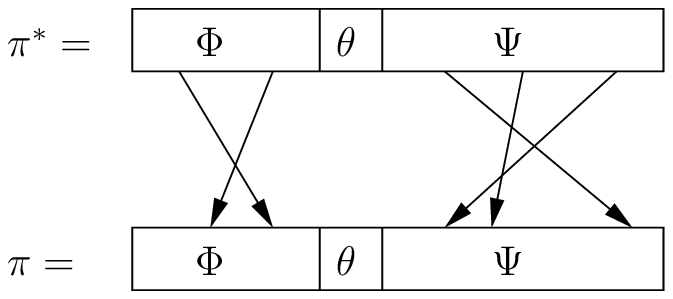}
\caption{Case 1. Permutation in $\Phi$ and $\Psi$.}\label{fig:case1}
\end{figure}

\textbf{Case 2.} \emph{Moving a set of users from $\Psi$ to the set
$\Phi$:} Assume a set $\Upsilon$ of users, $\Upsilon \subset \Psi$,
is moved from  $\Psi$ to the set $\Phi$ to generate a new
permutation $\pi$ (see Fig. \ref{fig:case2}). The rate of the user
$\theta$ in the new permutation is equal to:
\begin{equation}\label{eq:rate_teta2}
v_{\theta}(\pi)=f(\Phi \cup \Upsilon \cup \{\theta \}) - f(\Phi \cup
\Upsilon).
\end{equation}
From (\ref{poly:subm}), we can show that
\begin{equation}\label{ineq:pi_pib1}
 f(\Phi \cup \{\theta
\}) + f(\Phi \cup \Upsilon) \geq f(\Phi \cup \Upsilon \cup \{\theta
\})+f(\Phi).
\end{equation}
Using (\ref{eq:rate_pil}), (\ref{eq:rate_teta2}), and
(\ref{ineq:pi_pib1}), we conclude that $v_{\theta}(\pi) \leq
v_{\theta}(\pi^*)$, and therefore, $\min_i v_{\pi(i)}(\pi) \leq
\min_{i}v_{\pi^*(i)}(\pi^*)$.
\begin{figure}[tbhp]
\centering
\includegraphics[scale=1]{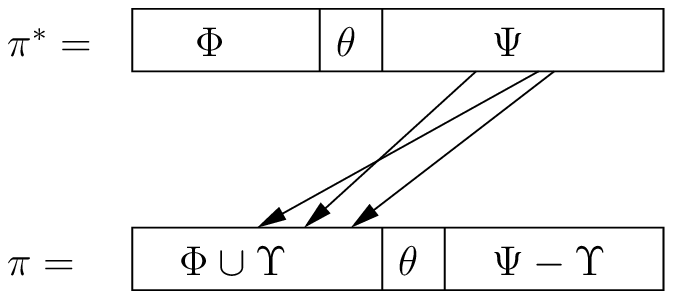}
\caption{Case 2. Moving a set of users from $\Psi$ to the set
$\Phi$.}\label{fig:case2}
\end{figure}

\textbf{Case 3.} \emph{Moving one or more users from the set $\Phi$
to the set $\Psi$ (with or without  moving some users from the set
$\Psi$ to the set $\Phi$):} Assume that one or more users move from
$\Phi$ to $\Psi$ (with or without  moving some users from the set
$\Psi$ to the set $\Phi$) to generate the new permutation $\pi$. As
depicted in Fig. \ref{fig:case3}, assume that the user $\nu$ is
positioned last  in the permutation $\pi$ among the users moved from
$\Phi$ to $\Psi$ (user $\pi(1)$ is positioned first and user
$\pi(a)$ is positioned last in the permutation $\pi$).
\begin{figure}[tbhp]
\centering
\includegraphics[scale=1]{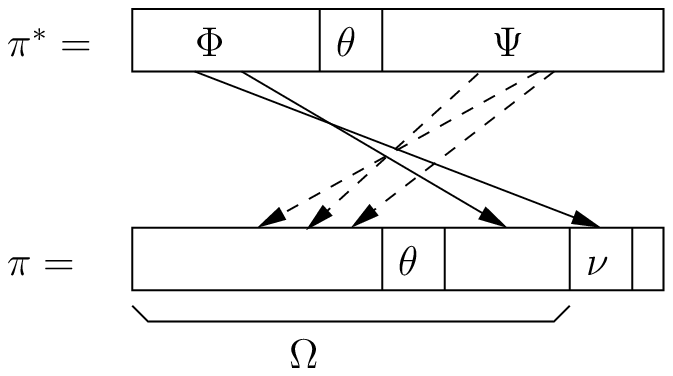}
\caption{Case 3. Moving one or more users from the set $\Phi$ to the
set $\Psi$ (with or without  moving some users from the set $\Psi$
to the set $\Phi$).}\label{fig:case3}
\end{figure}

Let $\Omega$ be the set of users located before the user $\nu$ in
the permutation $\pi$. Using (\ref{eq:v_pi}), we have,
\begin{equation}\label{eq:rate_vnu}
v_{\nu}(\pi)= f(\Omega \cup \{\nu \}) - f(\Omega).
\end{equation}
It is clear that,
\begin{equation}\label{eq:3}
\{\theta\} \cup \Phi   - \{\nu \} \subset\Omega.
\end{equation}
Using (\ref{poly:subm}) with $S=\Phi \cup \{\theta\}$ and
$T=\Omega$, and regarding (\ref{eq:3}), we have,
\begin{equation}\label{ineq:subm2}
f(\Omega \cup \{\nu \}) - f(\Omega) \leq f(\Phi \cup \{\theta\} ) -
f(\Phi \cup \{ \theta\}    - \{\nu \}).
\end{equation}
On the other hand, the user $\nu$ is in the set $\Phi$ in
permutation
 $\pi^*$. It means that in Step 2 of
 the algorithm, this user has been compared with other users in the set $\Phi \cup \{ \theta \}$  to
be located in the position $l$, but the user $ \theta$ has been
chosen for the position, i.e. $f\left(\Phi \cup
\{\theta\}-\{\theta\}\right) \leq f\left(\Phi \cup
\{\theta\}-\{\nu\}\right)$, therefore,
\begin{eqnarray}\label{ineq:4}
f\left(\Phi \right) \leq f\left(\Phi \cup \{\theta\}-\{\nu\}\right).
\end{eqnarray}

Using (\ref{eq:rate_pil}), (\ref{eq:rate_vnu}), (\ref{ineq:subm2}),
and (\ref{ineq:4}), we conclude that $v_{\nu}(\pi) \leq
v_{\theta}(\pi^*)$, and therefore, we have $\min_i v_{\pi(i)}(\pi)
 \leq \min_{i}v_{\pi^*(i)}(\pi^*)$. Note that the permutation of users
located before (or after) the user $\nu$ in the permutation $\pi$
does not increase $v_{\nu}(\pi)$.
\end{proof}

\textbf{Remark}: For multiple access channels,  the above algorithm
suggests that to attain the fairest corner point with successive
decoding, at each step, one should decode the strongest user (the
user with the highest rate, while the signals of the remaining users
are considered as interference). Note that in MAC, the corner point
corresponding to the specific permutation $\pi$ is obtained by the
successive decoding in the reverse order of the permutation.

It is worth mentioning that by using a similar algorithm, one can
find the corner point for which the maximum rate is minimum. The
algorithm is as follows:

\textbf{Algorithm II}
\begin{enumerate}
\item Set $\alpha =1$, $S= \varnothing$.
\item Set $\pi^*(\alpha)$ as
\begin{equation}\label{eq:1}
\pi^*(\alpha)= \arg \max_{z \in E , z \notin \mathcal{S}}
f\left(S+\{z\}\right).
\end{equation}
\item If $\alpha <a$, then  $S
\longleftarrow S \cup \{ \pi^*(\alpha) \}$, $\alpha\longleftarrow
\alpha+1$, and go to Step 2; otherwise stop.
\end{enumerate}

The optimality of the above algorithm can be proven by a similar
method as used to prove Theorem \ref{thm:max-min}.

\section{Optimal Rate-Vector on the Sum-Capacity Facet}\label{secIV}

\subsection{Max-Min Operation over a Polymatroid}

\begin{figure}[h]
  \hfill
  \begin{minipage}[t]{.45\textwidth}
    \begin{center}
      \epsfig{file=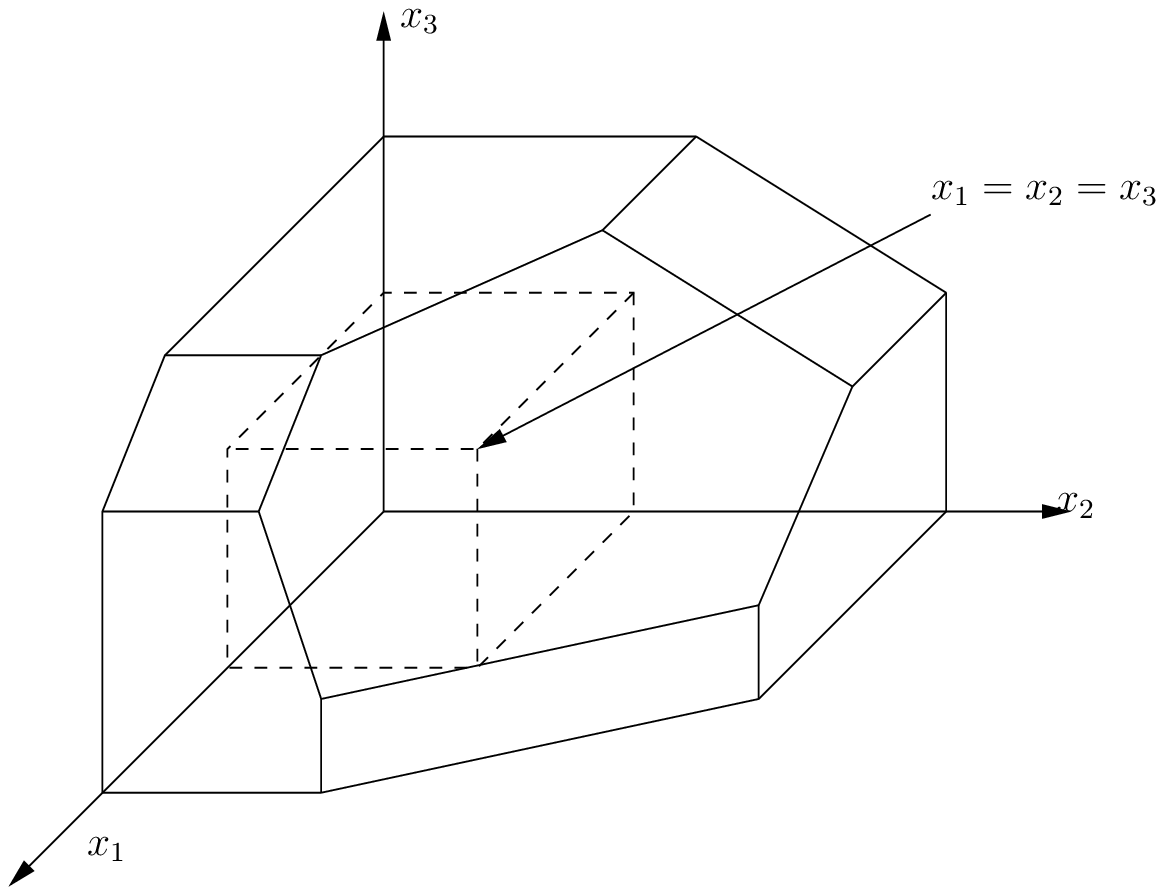, scale=0.7}
      \caption{All-Equal Rate-Vector Is on the Sum-Capacity Facet}
      \label{poly31}
    \end{center}
  \end{minipage}
  \hfill
  \begin{minipage}[t]{.45\textwidth}
    \begin{center}
      \epsfig{file=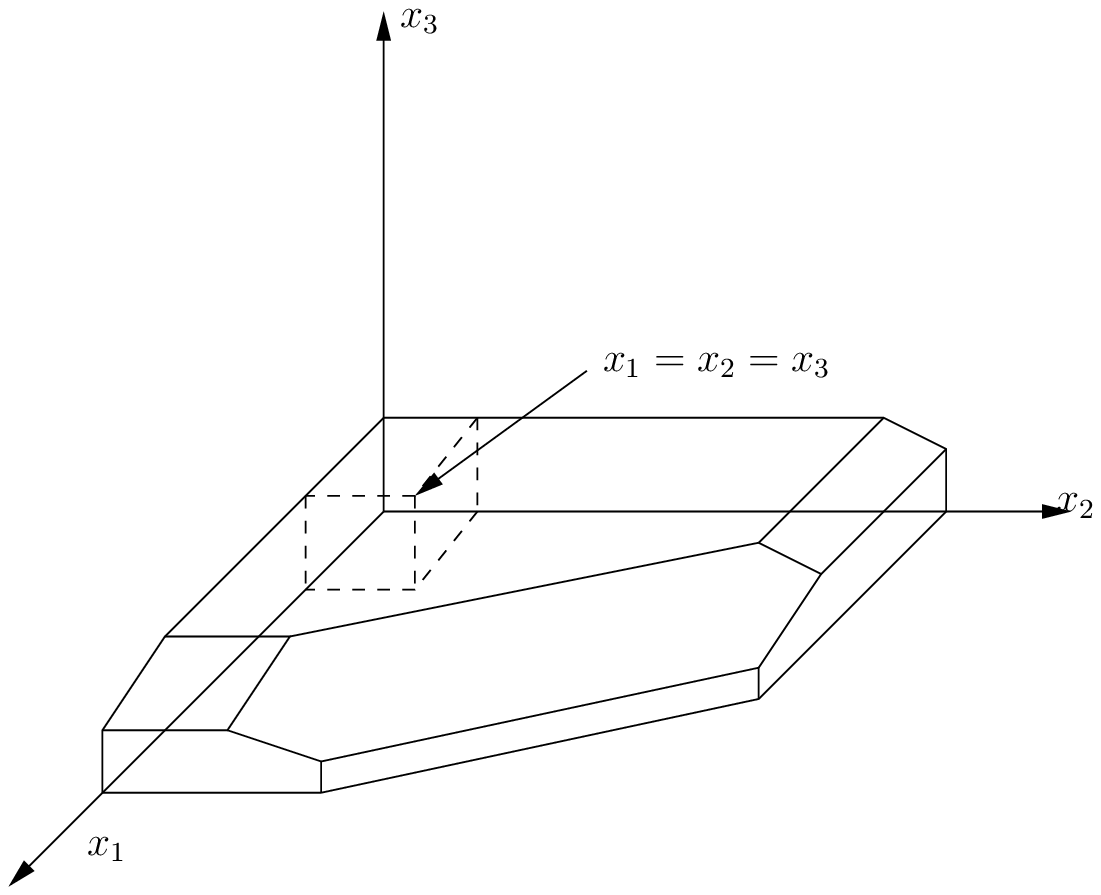, scale=0.7}
      \caption{All-Equal Rate-Vector Is NOT on the Sum-Capacity Facet}
      \label{poly32}
    \end{center}
  \end{minipage}
  \hfill
\end{figure}

In the following, the polymatroid properties are exploited to locate
an optimal fair point on the sum-capacity facet. For an optimal fair
point, the minimum rate among all the users should be maximized
(max-min rate). For a sum-capacity of $r_{\textrm{SC}}$, a fair rate
allocation would ideally achieve an equal rate of
$\dfrac{r_{\textrm{SC}}}{a}$ for the $a$ active users. Although this
rate-vector is feasible for some special cases (see Fig.
\ref{poly31}), it is not attainable in the general case (see Fig.
\ref{poly32}). The maximum possible value for the minimum entry of a
vector $\mathbf{x}$, where $\mathbf{x} \in \mathcal{B}(f,E)$, can be
computed using the following lemma.

\begin{lem}\label{lem:max_min}
In the polymatroid $\mathcal{B}(f,E)$, define
\begin{eqnarray}
\nonumber \delta&=&\max \ \min_{i \in  E } \ { x_i }. \\
&s.t.& (x_1,\ldots, x_a) \in \mathcal{B}(f,E).
\end{eqnarray}
Then,
\begin{equation}\label{eq:delta}
\delta=\min_{S \subset E, S \neq \emptyset }\frac{f(S)}{ |S|}.
\end{equation}
\end{lem}
\begin{proof}
Consider $\mathbf{x}\in \mathcal{B}(f,E)$, and let
$\sigma=\min_i{x_i}$. Therefore,
\begin{equation}
\forall S \subset E, \ \sigma|S|\leq \mathbf{x}(S).
\end{equation}
Noting $\forall S \subset E, \  \mathbf{x}(S) \leq f(S)$ and using
the above inequality, we have
\begin{equation}
\forall S \subset E, \ \sigma|S|\leq f(S).
\end{equation}
Consequently, $\sigma\leq \min_{S \subset E, \ S \neq
\emptyset}\frac{f(S)}{|S|}$. Therefore, $\min_{S \subset E, \ S \neq
\emptyset}\frac{f(S)}{|S|}$  provides an upper bound on $\min_i
x_i$. By selecting $\mathbf{x}=\delta \mathbf{1}_a \in
\mathcal{B}(f,E)$, where $\delta=\min_{S \subset E, \ S \neq
\emptyset}\frac{f(S)}{|S|}$, the upper bound is achieved, and the
proof is completed.
\end{proof}

In minimization (\ref{eq:delta}), if the minimizer is not the set
$E$, then $\delta$ (the optimal max-min value) is less than
$\frac{r_{\textrm{SC}}}{a}$ ( $r_{\textrm{SC}}=f(E)$ is the
sum-capacity),  and therefore, the ideal fairness is not feasible.
For example, in the polymatroid depicted in Fig \ref{poly32}, the
minimizing set in (\ref{eq:delta}) is the set $\{3\}$, and therefore
$\delta=f(\{3\})$.

In the following, a recursive algorithm is proposed to locate a rate
vector $\mathbf{x}^*$ on the sum-capacity facet which not only
attains the optimal max-min value $\delta$, but also provides
fairness among the users which have the rates higher than $\delta$.
The proposed algorithm partitions the set of active users into $t+1$
disjoint subsets, $S^{(0)},...,S^{(t)}$, such that in the $i$'th
subset the rate of all users is equal to $m^{(i)}, i=0,\cdots,t$,
where $\delta=m^{(0)}<m^{(1)}<\cdots<m^{(t)}$. Starting from
$m^{(0)}$, the algorithm maximizes $m^{(i)}$, $i=1,\cdots,t$, given
that $m^{(j)}$'s, $j=0,\cdots,i-1$, are already at their maximum
possible values.  To simplify this procedure, we establish a chain
of nested polymatroids, $\mathcal{B}(f^{(\alpha)},E^{(\alpha)})$,
$\alpha=0,\ldots,t$, where
\begin{equation}
\mathcal{B}(f^{(t)},E^{(t)}) \subset
\mathcal{B}(f^{(t-1)},E^{(t-1)})\subset \ldots \subset
\mathcal{B}(f^{(0)},E^{(0)})=\mathcal{B}(f,E).
\end{equation}

In this algorithm, we use the result of the following lemma.
\begin{lem}\label{lem:red_f}
 Let $E=\{1,\ldots,a\}$ and  $A \subset E$, $A \neq E$. If the set function $f:
2^E\longrightarrow \mathcal{R}_{+} $ is a rank function, then
$h~:~2^{E-A} \longrightarrow \mathcal{R}_{+}$, defined as
\begin{equation}
h(S)=f(S \cup A)-f(A), \quad S \subset E-A,
\end{equation}
is a rank function.
\end{lem}
\begin{proof}
By direct verification.
\end{proof}

Using the following algorithm, one can compute the rate-vector
$\mathbf{x}^*$.

\textbf{Algorithm III}
\begin{enumerate}
\item Initialize the iteration index $\alpha=0$, $E^{(0)}=E$, and
$f^{(0)}=f$. \item Find $m^{(\alpha)}$, where
\begin{equation}
 m^{(\alpha)}=\min_{S \subset E^{(\alpha)}, S \neq \emptyset
}\frac{f^{(\alpha)}(S)}{ |S|}.
\end{equation}
Set $S^{(\alpha)}$ equal to the optimizing subset.

\item For all $i
\in S^{(\alpha)}$, set $x^*_i=m^{(\alpha)}$.
\item Define the polymatroid
$\mathcal{B}(f^{(\alpha+1)},E^{(\alpha+1)})$, where
\begin{equation}
E^{(\alpha+1)}=E^{(\alpha)}-S^{(\alpha)},
\end{equation}
and $\forall S \subset E^{(\alpha+1)}$,
\begin{equation}\label{eq:rec_fa}
 \quad f^{(\alpha+1)}(S)=f^{(\alpha)}(S
\cup S^{(\alpha)} )-f^{(\alpha)}(S^{(\alpha)}) .
\end{equation}
\item If $E^{(\alpha+1)}
\neq \emptyset$, set $\alpha \longleftarrow \alpha+1$ and move to
step 2, otherwise stop.
\end{enumerate}

This algorithm computes the optimization sets $S^{(\alpha)}$,
$\alpha~=~0,\cdots,t$ and their corresponding $m^{(\alpha)}$, where
$E=\bigcup^t_{j=0} S^{(j)}$ and $x_i^* \in \{m^{(0)}, \cdots,
m^{(t)}\}, i=1, \cdots, a$.

\begin{figure}[tbhp]
\centering
\includegraphics[scale=1]{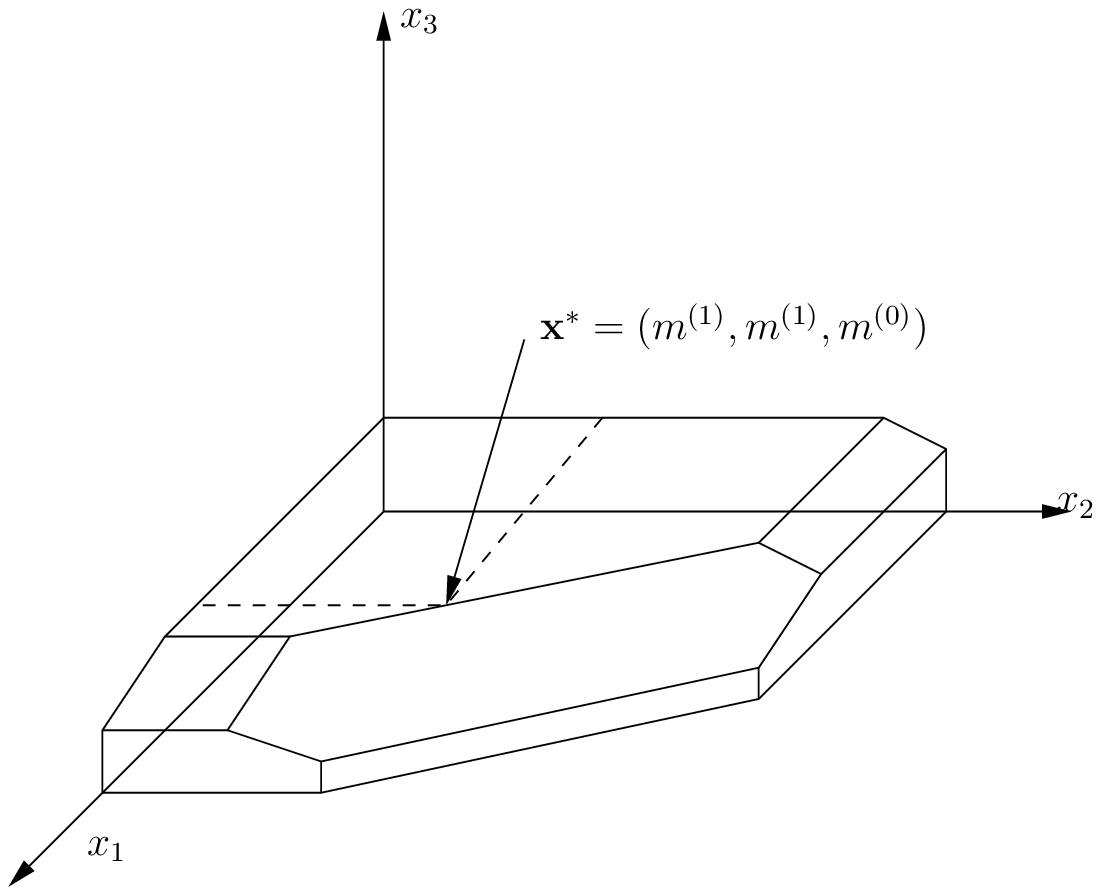}
\caption{The Fairest Rate Vector $\mathbf{x}^*$ on the Sum-Rate
Facet of the Polymatroid}\label{poly33}
\end{figure}

To provide better insight about the algorithm, let us apply it over
the polymatroids depicted in figures \ref{poly31}  and \ref{poly32}.
For the polymatroid in Fig. \ref{poly31}, the algorithm results in
$\mathbf{x}^*=(m^{(0)},m^{(0)},m^{(0)})$ where
$m^{(0)}=\frac{f(\{1,2,3\})}{3}$. For the polymatroid shown in Fig
\ref{poly32}, the resulting point is
$\mathbf{x}^*=(m^{(1)},m^{(1)},m^{(0)})$, where
$m^{(0)}=\frac{f(\{3\})}{1}$ and
$m^{(1)}=\frac{f^{(1)}(\{1,2\})}{2}=\frac{f(\{1,2,3\})-f(\{3\})}{2}$
(see Fig. \ref{poly33}).

In the following, we prove some properties of the  vector
$\mathbf{x}^*$.

\begin{thm}\label{thm:prop}
Assume that the algorithm III is applied over the polymatroid
$\mathcal{B}(f,E)$, then
\begin{enumerate}
\item[(I)]  $\mathbf{x}^* \in \mathcal{B}(f,E) $ and is located on
the sum-capacity facet $\mathbf{x}(E)= f(E)$.

\item[(II)] The minimum entry
of the vector $\mathbf{x}^*$ attains the optimum value determined by
Lemma \ref{lem:max_min} and
\begin{equation}
\delta=m^{(0)}<m^{(1)}<\cdots<m^{(t)}.
\end{equation}
\end{enumerate}
\end{thm}
\begin{proof}

\textbf{Part (I):} We show that $\mathbf{x}^* \in \mathcal{B}(f,E)$.
According to the algorithm, we have $m^{(0)}=\min_{S \subset E, S
\neq \emptyset }\frac{f(S)}{|S|}$, where $S^{(0)}$ is the minimizing
set. In addition, $x^*_i=m^{(0)}$ for all $i \in S^{(0)}$. It is
straight-forward to check that the assigned values for $x^*_i, i \in
S^{(0)}$, do not violate the constraints of the polymatroid
$\mathcal{B}(f,E)$, expressed in (\ref{ine:poly}). By substituting
the assigned values for $x_i, i \in S^{(0)}$, in the constraints of
the polymatroid $\mathcal{B}(f,E)$, the constraints over the
coordinate $i, \ i \in E-S^{(0)}$, are updated as follows: from the
definition of the polymatroid, we have a set of constraints on
$\mathbf{x}(S)$, $S \subset E-S^{(0)}$, which has the following
format:
\begin{equation}\label{eq:n_con}
\forall A \subset S^{(0)},  \mathbf{x}(S \cup A) \leq f^{(0)} (S
\cup A).
\end{equation}
Since $S \cap A=\emptyset$, then $\mathbf{x}(S \cup
A)=\mathbf{x}(S)+\mathbf{x}(A)$. Consequently, from
(\ref{eq:n_con}), we have,
\begin{equation}
\forall A \subset S^{(0)}, \mathbf{x}(S) \leq f^{(0)} (S \cup
A)-\mathbf{x}(A).
\end{equation}
Consequently, $\forall\  S \subset E-S^{(0)}$,
\begin{equation}
\mathbf{x}(S) \leq \min_{A \subset S^{(0)}} \{ f^{(0)} (S \cup
A)-\mathbf{x}(A)\}.
\end{equation}
We claim that $\min_{A \subset S^{(0)}} \{ f^{(0)} (S \cup
A)-\mathbf{x}(A)\}$ is equal to   $f^{(0)} (S \cup
S^{(0)})-f^{(0)}(S^{(0)})$. The proof is as follows:
\begin{eqnarray}
\forall A \subset S^{(0)},&& f^{(0)} (S \cup A)-\mathbf{x}(A) \\
&&\geq f^{(0)} (S \cup A)-f^{(0)}(A) \\
&&\geq f^{(0)} (S \cup S^{(0)})-f^{(0)}(S^{(0)}).
\end{eqnarray}
The first inequality relies on  the fact that $\forall A, \
\mathbf{x}(A)\leq f^{(0)}(A)$. The second inequality is proven by
using (\ref{poly:subm}) and the fact that $A \subset S^{(0)}$ and $S
\cap S^{(0)}=\emptyset$. It is easy to check that the above
inequalities change to equalities for $A=S^{(0)}$.

Regarding the above statements, for the non-allocated entries of
$\mathbf{x}$, we have the following set of constraints,
\begin{equation}\label{set:cond}
\forall S \subset E-S^{(0)}, \ \mathbf{x}(S) \leq f^{(0)}(S\cup
S^{(0)}) -f^{(0)}(S^{(0)}).
\end{equation}

Let us define $E^{(1)}=E^{(0)}-S^{(0)}$, $f^{(1)}(S)=f^{(0)}(S\cup
S^{(0)}) -f^{(0)}(S^{(0)})$, $\forall S \subset E^{(1)}$. By using
Lemma {\ref{lem:red_f}}, the set of constraints (\ref{set:cond}) on
$E^{(1)}$ defines the polymatroid $\mathcal{B}(f^{(1)},E^{(1)})$,
which is a subset of $\mathcal{B}(f, E)$. Now, we use the same
procedure that is applied for $\mathcal{B}(f^{(0)}, E^{(0)})$ over
$\mathcal{B}(f^{(1)}, E^{(1)})$, and continue recursively.
Therefore, in iteration indexed by $\alpha$, $\alpha=0,\ldots,t$,
the rates of a subset of coordinates are determined such that the
constraints of the polymatroid $\mathcal{B}(f^{(\alpha)},
E^{(\alpha)})$ are not violated. Since $\mathcal{B}(f^{(\alpha)},
E^{(\alpha)}) \subset \mathcal{B}(f, E)$, then $\mathbf{x}^* \in
\mathcal{B}(f, E)$. Direct verification proves that
$\mathbf{x}^*(E)=f(E)$.

\textbf{Part (II):} We must show that the smallest entries of
$\mathbf{x}^*$ is equal to $\min_{S \subset E}\frac{f(S)}{|S|}$.
According to the algorithm, for all $i \in E$, we have $x^*_{i} \in
\{m^{(0)},...,m^{(t)} \}$. Furthermore, $m^{(0)}=\min_{S \subset
E}\frac{f(S)}{|S|}$.

From the algorithm, we have
\begin{equation}\label{L0}
m^{(j)}= \frac{f^{(j)}(S^{(j)})}{|S^{(j)}|}=\min_{S \subset E^{(j)}}
\frac{f^{(j)}(S)}{|S|} < \frac{f^{(j)}( S^{(j+1)} \cup S^{(j)} )}{|
S^{(j+1)} \cup S^{(j)} |}=\frac{f^{(j)}(S^{(j+1)} \cup
S^{(j)})}{|S^{(j+1)}| + |S^{(j)}|}.
\end{equation}
Therefore,
\begin{eqnarray}
\label{L1} & m^{(j)} < \frac{f^{(j)}(S^{(j+1)} \cup
S^{(j)})}{|S^{(j+1)}| + |S^{(j)}|} & {\Longrightarrow}   \\
\label{L2}
 & m^{(j)} < \frac{f^{(j)}(S^{(j+1)} \cup
S^{(j)})-m^{(j)}|S^{(j)}|}{|S^{(j+1)}|} & \Longrightarrow  \\
\label{L3}& m^{(j)} < \frac{f^{(j)}(S^{(j+1)} \cup S^{(j)})-f^{(j)}(
S^{(j)})}{|S^{(j+1)}|}&=m^{(j+1)},
\end{eqnarray}
where (\ref{L3}) relies on LHS of (\ref{L0}). Consequently,  $
m^{(0)} < m^{(1)} < \ldots < m^{(t)} $ and the proof is complete.
\end{proof}

The remaining issue in Algorithm III is how to compute $\min_{S
\subset E, S \neq \emptyset }\frac{f(S)}{|S|}$. These types of
 problems are known as geometric minimizations.
In order to find the minimizer, the smallest value of $\beta$ is
desirable such that there is a set $S$ with $f(S)=\beta|S|$. For the
special case of single antenna Gaussian multiaccess channels,
computing such $\beta$ is very simple. For the general case, $\beta$
can be computed by Dinkelbach's discrete Newton method as follows
\cite{FleIwa03}.

The algorithm is initialized by setting $\beta$ equal to $f(E)/|E|$,
which is an upper bound for optimum $\beta$. Then, a minimizer $Y$
of $f(S)-\beta|S|$ is calculated, as will be explained later. Since
$f(E)-\beta|E|= 0$, then $f(Y)-\beta|Y|\leq 0$. If $f(Y)-\beta|Y|=
0$, the current $\beta$ is optimum. If $f(Y)-\beta|Y|< 0$, then we
update $\beta=f(Y)/|Y|$, which provides an improved upper bound. By
repeating this operation, the optimal value of $\beta$ will
eventually be calculated~\cite{FleIwa03}. It is shown that the
number of $\beta$ visited by the algorithm is at most $|E|$
\cite{FleIwa03}.

Using this approach, the minimization problem $\min_{S \subset E, S
\neq \emptyset }\frac{f(S)}{|S|}$ is changed to $\min_{S \subset E,
S \neq \emptyset } f(S)-\beta |S|$. By direct verification of
(\ref{poly:subm}), it is easy to see that $f(S)-\beta |S|$ is a
submodular function. There have been a lot of research on submodular
minimization problems \cite{FleIwa03, Sch00, IwFlFu01}. In
\cite{Sch00, IwFlFu01}, the first combinatorial polynomial-time
algorithms for solving submodular minimization problems are
developed. These algorithms design a  strongly polynomial
combinatorial algorithm for testing membership in polymatroid
polyhedra.

%
%
\subsection{Decomposition of the Time-Sharing Problem}

In the following, we take advantage of the special properties of
$\mathbf{x}^*$ and polymatroids to break down the time-sharing
problem to some lower dimensional subproblems. In the previous
subsection, a chain of nested polymatroids
$\mathcal{B}(f^{(\alpha)},E^{(\alpha)})$, $\alpha=0,\ldots,t$, is
introduced, where $\mathcal{B}(f^{(\alpha-1)},E^{(\alpha-1)})
\subset \mathcal{B}(f^{(\alpha)},E^{(\alpha)})$ for
$\alpha=1,\ldots,t$. Since $S^{(j)} \subset E^{(j)}$ for
$j=0,\ldots, t$ and regarding the definition of polymatroid,
$\mathcal{B}(f^{(j)},S^{(j)})$, $j=1,\ldots,t$, is a polymatroid,
which is defined on the dimensions $S^{(j)}$.  According to the
proof of Theorem \ref{thm:prop},  the vector
$m^{(j)}\mathbf{1}_{|S^{(j)}|} \in \mathcal{B}(f^{(j)},S^{(j)})$ is
on the hyperplane $\mathbf{x}(S^{(j)})=f(S^{(j)})$. Let
$\{\pi^{(j)}_{\gamma_j}, \gamma_j=1,\ldots, |S^{(j)}|!\}$ be the set
of all permutations of the set $S^{(j)}$, and
$\mathbf{u}^{(j)}(\pi^{(j)}_{\gamma_j})$ be the corner point
corresponding to the permutation $\pi^{(j)}_{\gamma_j}$ in the
polymatroid $\mathcal{B}(f^{(j)},S^{(j)})$. Then, there exist the
coefficients $ 0 \leq \lambda_{\gamma_j}^{(j)} \leq 1$,
$\gamma_j=1,\ldots,|S^{(j)}|!$, such that
\begin{equation}\label{eq:part_time}
m^{(j)}\mathbf{1}_{|S^{(j)}|}=
\sum_{\gamma_j=1}^{|S^{(j)}|!}\lambda_{\gamma_j}^{(j)}
\mathbf{u}^{(j)} \big{(}\pi^{(j)}_{\gamma_j} \big{)},
\end{equation}
where
\begin{equation}
 \sum_{\gamma_j=1}^{|S^{(j)}|!}\lambda_{\gamma_j}^{(j)} =1.
\end{equation}
Note that $E=\bigcup_{j=0}^t S^{(j)}$. Consider a permutation
$\pi^{(j)}_{\gamma_j}$ as one of the total $|S^{(j)}|!$ permutations
of $S^{(j)}$, for $j=0,\cdots,t$, then the permutation $\pi$ formed
by concatenating these permutations, i.e. $\pi =
\left(\pi^{(t)}_{\gamma_t} , \cdots , \pi^{(0)}_{\gamma_0}\right)$,
is a permutation on the set $E$.

\begin{thm}
Consider the permutation $\pi = \left( \pi^{(t)}_{\gamma_t} , \cdots
, \pi^{(0)}_{\gamma_0}\right)$ of the set $E$.
\begin{itemize}
\item[(I)] The corner point corresponding to the permutation $\pi$ in the
polymatroid $\mathcal{B}(f,E)$ is
\begin{equation}
v_{i}(\pi)=u^{(j)}_i(\pi^{(j)}_{\gamma_j}), \quad \textup{for} \quad
i \in S^{(j)},
\end{equation}
where $\mathbf{u}^{(j)}(\pi^{(j)}_{\gamma_j})$ is the corner point
of the polymatroid $\mathcal{B}(f^{(j)},S^{(j)})$ corresponding to
the permutation $\pi^{(j)}_{\gamma_j}$, and
${u}^{(j)}_i(\pi^{(j)}_{\gamma_j})$ denotes the value of
$\mathbf{u}^{(j)}(\pi^{(j)}_{\gamma_j})$ over the dimension $i$, $i
\in S^{(j)}$.

\item[(II)] The vector $\mathbf{x}^*$ is in the convex hull of the
set of corner points corresponding to the following set of
permutations
\begin{equation}
\left \{ \left( \pi^{(t)}_{\gamma_t} , \cdots ,
\pi^{(0)}_{\gamma_0}\right), 1 \leq\gamma_{t} \leq |S^{(t)}|! ,
\ldots,  1 \leq\gamma_0 \leq |S^{(0)}|! \right \},
\end{equation}
where the coefficient of the corner point corresponding to the
permutation $\pi = \left( \pi^{(t)}_{\gamma_t} , \cdots ,
\pi^{(0)}_{\gamma_0}\right)$ is equal to $\lambda_{\gamma_{t}}^{(t)}
\ldots \lambda_{\gamma_0}^{(0)}$, i.e.
\begin{equation}
\mathbf{x}^*= \sum_{\gamma_t=1}^{|S^{(t)}|!} \ldots
\sum_{\gamma_0=1}^{|S^{(0)}|!} \lambda_{\gamma_t}^{(t)} \ldots
\lambda_{\gamma_0}^{(0)} \mathbf{v}\big{(}\left(
\pi^{(t)}_{\gamma_t} , \cdots , \pi^{(0)}_{\gamma_0}\right)\big{)}.
\end{equation}
\end{itemize}
\end{thm}
\begin{proof}
\textbf{Part (I)} From recursive equation (\ref{eq:rec_fa}), we can
show that
\begin{equation}\label{eq:f_expand}
\textrm{For} \  S \in E-\bigcup_{i=0}^{j-1}S^{(i)}, \quad
f^{(j)}(S)=f \left ( S \cup \left \{ \bigcup_{i=0}^{j-1}S^{(i)}
\right\} \right )-f \left (  \left \{ \bigcup_{i=0}^{j-1}S^{(i)}
\right\} \right ).
\end{equation}
Consider the permutation $\pi=\left( \pi^{(t)}_{\gamma_t} , \cdots ,
\pi^{(0)}_{\gamma_0}\right)$. Set $\xi = \sum_{i=1}^{j} |S^{(i)}| $.
By using (\ref{eq:v_pi}) and (\ref{eq:f_expand}), for $\xi < \kappa
\leq \xi+ |S^{(j+1)}|$, $v_{\pi(\kappa)}(\pi)$ is equal to
\begin{eqnarray}
v_{\pi(\kappa)}(\pi)= && f \left( \{ \pi(1),\ldots, \pi(\kappa) \}
\right) - f\left( \{ \pi(1),\ldots, \pi(\kappa-1)\}
\right) \\
= &&  f \left( \left \{ \bigcup_{i=0}^{j-1}S^{(i)},
\pi(\xi+1)\ldots, \pi(\kappa) \right \} \right) - f\left( \left \{
\bigcup_{i=0}^{j-1}S^{(i)},\pi(\xi+1),\ldots, \pi(\kappa-1) \right
\} \right)\\
=&& f^{(j)}\left( \left \{ \pi(\xi+1)\ldots, \pi(\kappa) \right \}
\right)-f^{(j)}\left( \left \{ \pi(\xi+1)\ldots, \pi(\kappa-1)
\right \} \right) \label{eq:RHS}.
\end{eqnarray}
According to definition of polymatroid and its corner points, the
RHS of (\ref{eq:RHS})  is the value of
$u^{(j)}_{\pi(\kappa)}(\pi^{(j)})$ in the corresponding corner point
of the polymatroid $\mathcal{B}(f^{(j)},S^{(j)})$.

\textbf{Part (II)} Since $\sum_{\gamma_0=1}^{|S^{(0)}|!}
\lambda_{\gamma_0}^{(0)}=1$ and by using (\ref{eq:part_time}) and
part (I) of the theorem, it is easy to verify that the $i^{th}$, $i
\in S^{(0)}$, entry of
\begin{equation}
\sum_{\gamma_0=1}^{|S^{(0)}|!} \lambda_{\gamma_0}^{(0)}
\mathbf{v}\big{(}\pi^{(t)}_{\gamma_t} ,\ldots,
\pi^{(0)}_{\gamma_0}\big{)}
\end{equation}
is equal to $m^{(0)}$. Similarly, the entry $i$, $i \in S^{(1)}$, of
\begin{equation}
\sum_{\gamma_1=1}^{|S^{(1)}|!}
\lambda_{\gamma_1}^{(1)}\sum_{\gamma_0=1}^{|S^{(0)}|!}
\lambda_{\gamma_0}^{(0)} \mathbf{v}\big{(}\pi^{(t)}_{\gamma_t}
,\ldots, \pi^{(0)}_{\gamma_0}\big{)},
\end{equation}
is equal to $m^{(1)}$, while the entry $i$, $i \in S^{(0)}$, remains
$m^{(0)}$. By continuing this procedure, part (II) of the algorithm
is proven.
\end{proof}

Regarding the above statements, the problem of finding time-sharing
coefficients is decomposed to some lower dimensional subproblems. In
each sub-problem, the objective is to find the coefficients of the
time-sharing among the corner points of the polymatroid
$\mathcal{B}(f^{(j)},S^{(j)})$, $j=0,\ldots,t$, to attain
$m^{(j)}\mathbf{1}_{|S^{(j)}|}$. In this part, we present an
algorithm which finds the coefficients of the time-sharing over the
corner points of a general polymatroid $\mathcal{B}(f,E)$ to attain
a vector $\mathbf{x}$ located on the face of the polymatroid.

\textbf{Algorithm IV}
\begin{enumerate}
\item Initialize $\alpha=1$, $\mathbf{u}_1=\mathbf{v}(\pi^*)$ (the
fairest corner point obtained by algorithm I).
\item Solve  the linear optimization problem
\begin{eqnarray}
\nonumber & \max \tau & \\
\nonumber s.t. & \sum_{i=1}^{\alpha} \mu_i \mathbf{u}_i -\mathbf{x} \geq  \tau &\\
& 0 \leq \mu_i \leq 1 &
\end{eqnarray}
Let $\mu^{\alpha}_i$, $i=1,\ldots,\alpha$ be the optimizing
coefficients.
\item If $\mathbf{x}=\sum_{i=1}^{\alpha} \mu^{\alpha}_i
\mathbf{u}_i$, Stop.
\item $\alpha \longleftarrow \alpha +1$. Set $\mathbf{e}=\mathbf{x}-\sum_{i=1}^{\alpha} \mu^{\alpha}_i
\mathbf{u}_i$ and determine the permutation $\pi$ for which
$\mathbf{e}_{\pi(1)} \geq \mathbf{e}_{\pi(2)} \geq \ldots \geq
\mathbf{e}_{\pi(|E|)}$. Set $\mathbf{u}_{\alpha}=\mathbf{v}(\pi)$
and move to step 2.
\end{enumerate}
The idea behind the algorithm is as follows. In each step, the
time-sharing among some corner points is performed. If the resulting
vector is equal to $\mathbf{x}$,  the answer is obtained; otherwise
a permutation $\pi$ is determined such that
 $\mathbf{e}_{\pi(1)} \geq
\mathbf{e}_{\pi(2)} \geq \ldots \geq \mathbf{e}_{\pi(|E|)}$, where
the error vector $\mathbf{e}$ represents the difference between the
vector $\mathbf{x}$ and resulting vector from time-sharing. We can
compensate the error vector $\mathbf{e}$ by including an appropriate
corner point in the set of corner points participating in
time-sharing. Clearly, the best one to be included is the one which
has the highest possible rate for user $\pi(1)$ and lowest possible
rate for user $\pi(|E|)$. Apparently, this corner point is
$\mathbf{v}(\pi)$, computed by
 algorithm IV.

Note that Algorithm IV can be applied over the sub-polymatroids
$\mathcal{B}(f^{(j)},S^{(j)})$, $j=0,\ldots,t$, to attain
$m^{(j)}\mathbf{1}_{|S^{(j)}|}$ or directly applied over the
original polymatroid to attain $\mathbf{x}^*$ . If $a$ and $|S^{j}|$
are relatively small numbers, the decomposition method has less
complexity, otherwise applying Algorithm IV over the original
problem is less complex.

\subsection{Decomposition of Rate-Splitting Approach}
As mentioned, an alternative approach to achieve any rate-vector on
the sum-capacity facet of MAC is {\em rate
splitting}~\cite{rate_spil1,rate_spil2}. This method is based on
splitting all input sources except one into two parts, and treating
each spilt input as two virtual inputs (or two virtual users). Thus,
there are at most $2a-1$ virtual users. It is proven that by
splitting the sources appropriately and successively decoding
virtual users in a suitable order, any point on the sum-capacity
facet can be attained.

Similar to the time-sharing part, we prove that to attain the rate
vector $\mathbf{x}^*$, the rate-splitting procedure can be
decomposed into some lower dimensional subproblems. Consider a MAC,
where the capacity region is represented by polymatroid
$\mathcal{B}(f,E)$ and the vector $\mathbf{x}^*$, derived in
Algorithm III, is on its face. Assume that the users in the set
$S^{(j)}$ are decoded before the set of users in
$\{S^{(j-1)},S^{(j-2)},\ldots,S^{(0)}\}$ and after the users in the
set $\{S^{(t)}, \ldots,S^{(j+2)} ,S^{(j+1)}\}$ .Therefore, by
similar discussion used in  (\ref{eq:n_con}) to (\ref{set:cond}), we
conclude that the rate of the users in the set $S^{(j)}$ is
characterized by the polymatroid $\mathcal{B}(f^{(j)},S^{(j)})$,
where the rate-vector $m^{(j)} \mathbf{1}_{|S^{(j)}|} $ is on its
face. Regarding the results presented
in~\cite{rate_spil1,rate_spil2}, we can attain the rate-vector
$m^{(j)} \mathbf{1}_{|S^{(j)}|}$ by properly splitting the sources
of all inputs, except for one, in the set $S^{(j)}$ to form
$2|S^{(j)}|-1$ virtual users and by choosing the proper order of the
decoding of the virtual users. Consequently, using  algorithm V
(below), we achieve the rate-vector $\mathbf{x}^*$ in the original
polymatroid.

\textbf{Algorithm V}
\begin{enumerate}
\item Apply rate-splitting approach to attain the rate-vector
$m^{(j)} \mathbf{1}_{|S^{(j)}|} $ on the face of the polymatroid
$\mathcal{B}(f^{(j)},S^{(j)})$, for $j=0,\ldots,t$. Therefore, for
each $j$, $0 \leq j \leq t$,  at most $2|S^{(j)}|-1$ virtual users
are specified with a specific order of decoding.

\item Starting from $j=t$, decode the virtual users in the set $S^{(j)}$ in the
order found in Step 1. Set $j \leftarrow j-1$. Follow the procedure
until $j <0$.
\end{enumerate}

\section{Conclusion}
We considered the problem of fairness for a class of systems for
which a subset of the capacity region forms a polymatroid structure.
The main purpose is to find a point on the sum-capacity facet which
satisfies a notion of fairness among active users. This problem is
addressed in cases where the complexity of achieving interior points
is not feasible, and where the complexity of achieving interior
points is feasible. For the first case, the corner point for which
the minimum rate of the active users is maximized (max-min corner
point) is desired for signaling. A simple greedy algorithm is
introduced to find the optimum max-min corner point. For the second
case, the polymatroid properties are exploited to locate a
rate-vector on the sum-capacity facet which is optimally fair in the
sense that the minimum rate among all users is maximized (max-min
rate). In the case that the rate of some users can not increase
further (attain the max-min value), the algorithm recursively
maximizes the minimum rate among the rest of the users. It is shown
that the problems of deriving the time-sharing coefficients and
rate-spitting scheme can be solved by decomposing the problem to
some lower-dimensional subproblems. In addition, a fast algorithm to
compute the time-sharing coefficients to attain a general point on
the sum-capacity facet is proposed.

\section*{Acknowledgement}
The authors would like to thank Mr. Mohammad H. Baligh and Mr.
Shahab Oveis Gharan for helpful discussions.

\small{

\bibliographystyle{IEEE}
}
\end{document}